\documentclass[11pt,oneside,a4paper]{article}

\usepackage{siunitx}
\usepackage{amsmath}
\usepackage{titling}
\usepackage{authblk}
\usepackage{graphicx}
\usepackage{hyperref}
\usepackage[a4paper,left=3cm,right=2cm,top=2.5cm,bottom=2.5cm]{geometry}

\title{Time-resolved photoemission electron microscopy on a ZnO surface using an extreme ultraviolet attosecond pulse pair}

\author[1,2,*]{Jan Vogelsang}
\author[2]{Lukas Wittenbecher}
\author[2]{Sara Mikaelsson}
\author[2]{Chen Guo}
\author[2]{Ivan Sytcevich}
\author[2]{Anne-Lise Viotti}
\author[2]{Cord L. Arnold}
\author[2]{Anne L'Huillier}
\author[2]{Anders Mikkelsen}

\affil[1]{Institute of Physics, University of Oldenburg, 26129 Oldenburg, Germany}
\affil[2]{Department of Physics, Lund University, 22100 Lund, Sweden}

\affil[*]{jan.vogelsang@uni-oldenburg.de}

\begin{document}

\begin{titlingpage}
\maketitle
\begin{abstract}
Electrons photoemitted by extreme ultraviolet attosecond pulses derive spatially from the first few atomic surface layers and energetically from the valence band and highest atomic orbitals. As a result, it is possible to probe the emission dynamics from a narrow two-dimensional region in the presence of optical fields as well as obtain elemental specific information. However, combining this with spatially-resolved imaging is a long-standing challenge because of the large inherent spectral width of attosecond pulses as well as the difficulty of making them at high repetition rates. Here we demonstrate an attosecond interferometry experiment on a zinc oxide (ZnO) surface using spatially and energetically resolved photoelectrons. We combine photoemission electron microscopy with near-infrared pump - extreme ultraviolet probe laser spectroscopy and resolve the instantaneous phase of an infrared field with high spatial resolution. Our results show how the core level states with low binding energy of ZnO are well suited to perform spatially resolved attosecond interferometry experiments. We observe a distinct phase shift of the attosecond beat signal across the laser focus which we attribute to wavefront differences between the pump and the probe fields at the surface. Our work demonstrates a clear pathway for attosecond interferometry with high spatial resolution at atomic scale surface regions opening up for a detailed understanding of nanometric light-matter interaction.
\end{abstract}
\end{titlingpage}

\section{Introduction}

Attosecond interferometry utilizes extreme ultraviolet (XUV) and soft x-ray pulses with photon energies of 10 to hundreds of eV and a duration of down to \SI{50}{as} to unravel ultrafast charge carrier dynamics in atoms, molecules, liquids and solids \cite{sainadh_attosecond_2019,nisoli_attosecond_2017,jordan_attosecond_2020,cavalieri_attosecond_2007}.
The exquisite temporal resolution allows a real time observation of charges interacting with the oscillating electric field of visible light – the fundamental limit of opto-electronic applications \cite{ossiander_speed_2022,sederberg_attosecond_2020,bionta_-chip_2021,sommer_attosecond_2016}.
Such dynamics are studied inside the bulk of a material for example via the attosecond pulse generation process itself or via transient absorption spectroscopy, both combined with theoretical modeling \cite{vampa_linking_2015,luu_extreme_2015,goulielmakis_high_2022,lucchini_attosecond_2016,schultze_attosecond_2014,kobayashi_direct_2019}.
On surfaces the interaction of photoelectrons (excited by XUV attosecond pulses) and strong optical fields has been used to unravel the fastest charge carrier dynamics at the material vacuum interface \cite{cavalieri_attosecond_2007,tao_direct_2016,chen_distinguishing_2017,siek_angular_2017}.
Interferometric methods employing no attosecond but two visible to infrared pulses with variable delay are able to probe either sub-cycle phase shifts or dynamics on the 10-fs scale and longer.
This concept has been successfully applied to photoemission electron microscopy (PEEM) experiments in a range of studies in particular on charge carrier dynamics in nanostructures and quasi-particle effects on structured surfaces \cite{spektor_revealing_2017, podbiel_imaging_2017, dai_plasmonic_2020, zhong_nonlinear_2020, wittenbecher_unraveling_2021}
Extending methods like PEEM to using attosecond XUV pulses requires fundamentally new developments with their respective challenges.
In particular, the combination of PEEM with attosecond XUV pulses requires the simultaneous spectroscopic analysis of, compared to optical experiments, highly energetic photoelectrons and hence significantly better statistics while the optical pulse source is less stable than traditional laser systems and, until recently, not available at high repetition rates above \SI{100}{kHz}.
Furthermore, attosecond techniques often require strong optical fields to measurably alter an electron's trajectory -- a problem since strong fields often go along with strong photoelectron emission while PEEM experiments are limited to less than one emitted electron per laser shot due to Coulomb repulsion.
A consequence has been that while PEEM using attosecond XUV pulses is in principle promising for ultrafast studies \cite{stockman_attosecond_2007}, it has been difficult to realize in practice beyond initial imaging experiments \cite{mikkelsen_photoemission_2009} already more than 10 years ago.

Still, the prospect of nanoscale near-field sampling attracts a lot of interest due to its potentially widespread applications.
On a solid surface the possibility for microscopic structuring \cite{gates_new_2005,hengsteler_bringing_2021} allows the manipulation of local optical fields \cite{aeschlimann_adaptive_2007,novotny_principles_2006}.
Local fields in turn have important applications in for example charge carrier excitation \cite{boolakee_light-field_2022}.
The influence of shaping and enhancing local optical fields has been studied in recent years using different ultrafast methods – for example ultrafast electron microscopy \cite{vogelsang_coherent_2021,feist_ultrafast_2017}, ultrafast scanning tunneling microscopy \cite{garg_real-space_2021,cocker_tracking_2016} or ultrafast scanning near-field optical microscopy \cite{eisele_ultrafast_2014}.
While recent studies have shown local near field electron dynamics with nanoscale resolution \cite{gaida_attosecond_2023, nabben_attosecond_2023}, these are transmission experiments which do not directly probe the lateral variations close to the surface, which are crucial for their functionalization e.g. in 2D materials.
Attosecond interferometry offers the required temporal resolution and surface sensitivity but so far lacks the lateral spatial resolution to study individual nanostructures.
Only few attosecond time-resolved experiments on nanostructures without nanoscale resolution have been demonstrated \cite{forg_attosecond_2016, zherebtsov_controlled_2011,seiffert_strong-field_2022}.

When quantitative information about nanoscale fields are required, the probing optical field inside the laser focus needs to be known both temporally and in particular spatially – a fact so far largely ignored for spatially averaging attosecond experiments. Recently, it was shown that the phase evolution inside an optical focus of an ultrashort light pulse is far from trivial \cite{hoff_tracing_2017} as it is expected for attosecond light pulses \cite{wikmark_spatiotemporal_2019} and can be exploited for e.g. controlled high-order harmonic generation \cite{vincenti_attosecond_2012}. While measurements of the wavefront of an attosecond pulse have been demonstrated \cite{lee_wave-front_2003,dacasa_single-shot_2019}, the impact on a time-resolved experiment with spatial resolution has not been studied yet. The additional spatial resolution requires significantly more statistics of single electron events.

Here, we demonstrate an attosecond interferometry experiment imaging photoelectrons from a surface using PEEM. This is achieved by carefully preparing a clean ZnO crystal surface for photoemission of electrons from well-defined energetic states of both zinc (Zn) and oxygen (O) atoms. Electrons are emitted using pairs of extreme ultraviolet (XUV) attosecond pulses and subsequently accelerated near the surface using a time-delayed few-cycle near-infrared light field. We detect the electron emission site spatially resolved in a photoemission electron microscope (PEEM) equipped with an imaging energy filter. By analyzing the kinetic energy of the photoelectrons, we observe a distinct acceleration and deceleration of photoelectrons from Zn-3d and O-2p states as they interact with the infrared laser field at the surface of the ZnO crystal. The spatially-resolved image of the dynamics for varying delays between near-infrared (NIR) and XUV allows us to extract information about the relative phase of the ultrashort pulses involved and hence image their dynamics in the near-field of a surface. While the PEEM allows imaging with a spatial resolution of approximately \SI{20}{nm} \cite{lin_time_2009}, taking advantage of this will require careful treatment of the observed phase differences to allow a quantitative comparison of field dynamics on a heterogeneous surface. Our findings are an important step towards utilizing attosecond XUV pulses in a time-resolved photoemission electron microscope with high spatial resolution at the atomic scale region perpendicular to the surface.

\section{Results and discussion}
\subsection{Experimental setup}

\begin{figure}[h!tbp]
\centering
\includegraphics[width=\linewidth]{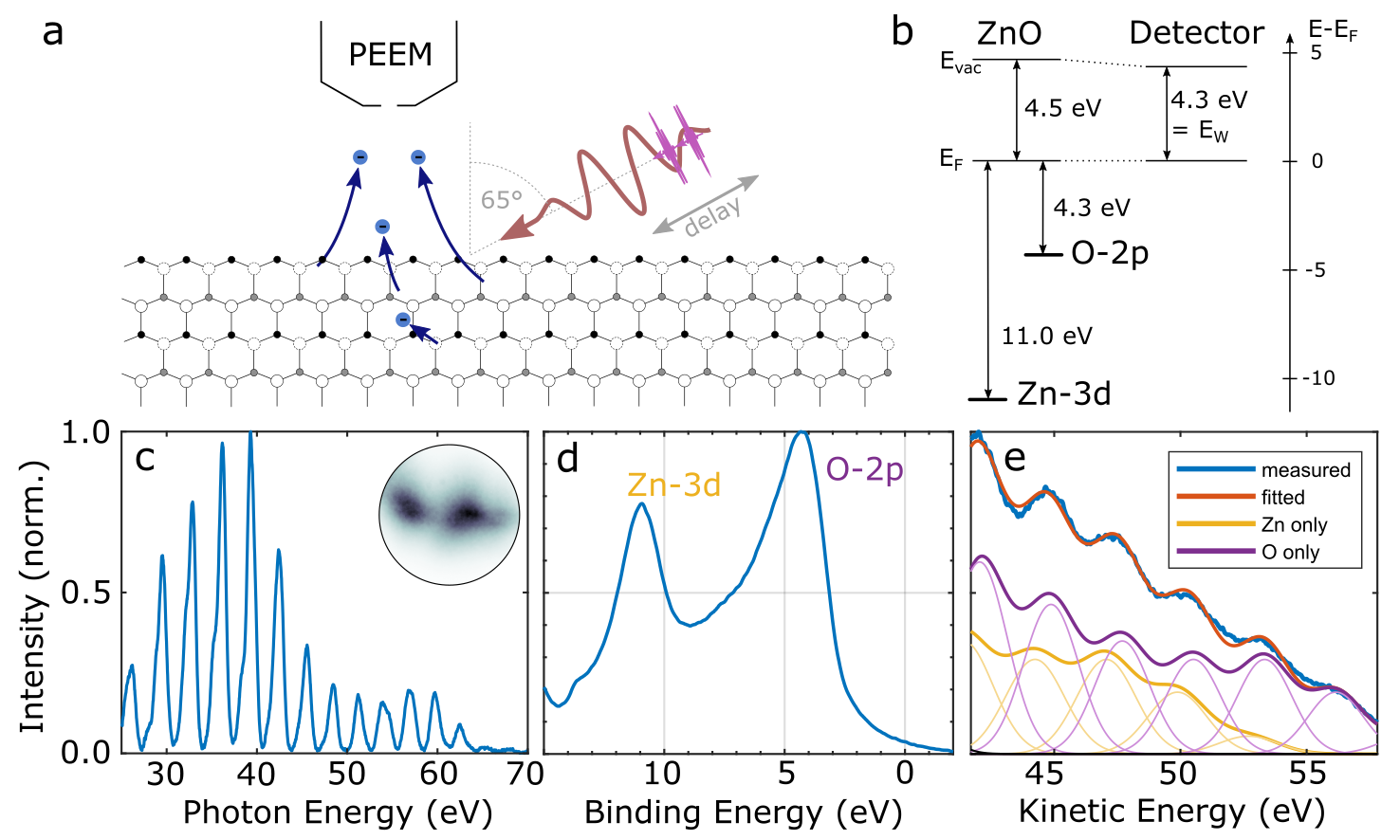}
\caption{Characterization of the experimental setup. a) Schematic of the steps involved in the experiment. A pair of XUV pulses (drawn in violet) photoemits electrons from a ZnO crystal. The electrons experience the dynamic field of an NIR laser pulse (drawn in red) close to the surface at a variable waiting time. The emission site of the electrons as well as their kinetic energy after interaction with the NIR field are recorded using a photoemission electron microscope (PEEM). b) Energy diagram of the ZnO surface and the electron detector which are electrically contacted and thus have their Fermi levels aligned. c) Optical spectrum of the XUV pulses used for photoemitting electrons from the surface. The inset shows the linear photoemission pattern generated by the XUV pulses from a ZnO surface. The field of view (FOV) of the inset is \SI{180}{\micro m}. d) Measurement of the electronic states close to the Fermi level of the ZnO surface. It was performed using a helium gas discharge lamp emitting a photon energy of \SI{21.2}{eV} and a hemispherical analyzer for electron detection after photoemission. e) Kinetic energy spectrum of photoelectrons emitted from a ZnO surface using the spectrum shown in c). The energy-dependent emission cross section of the Zn-3d and O-2p states indicated in d) was used as a fitting parameter in combination with the optical spectrum shown in c) to replicate the modulated spectrum shown in blue. The contribution to the emission from Zn-3d and O-2p by the individual harmonics is shown in lighter colors, respectively.}
\label{fig:Fig1}
\end{figure}

We utilize few-cycle light pulses in the near-infrared spectral region, in the following called NIR pulses, with a central wavelength of $\lambda=\SI{850}{nm}$, a duration of \SI{6.4}{fs} (full width at half the maximum of the intensity envelope) and a stable carrier to envelope phase (CEP; \SI{160}{mrad} rms single shot) delivered by an optical parametric chirped-pulse amplifier laser system at a repetition rate of \SI{200}{kHz} with an average power of \SI{3.0}{W}. \SI{80}{\percent} of the laser pulse energy (\SI{12}{\micro J}) is used to generate high order harmonics with photon energies up to \SI{70}{eV} (Figure \ref{fig:Fig1}c) in a high density ($\approx$ \SI{5}{bar}) argon gas jet \cite{mikaelsson_high-repetition_2020}. The CEP of the few-cycle NIR pulses is adjusted so that mainly two half cycles of the optical field contribute to the HHG process, resulting in the generation of two attosecond pulses. As a result, after separation from the NIR pulses via a \SI{200}{nm} thin aluminum filter, pairs of attosecond pulses are available with a temporal separation of half a cycle of the pump pulses ($\approx$ \SI{1.4}{fs}).

The pulse energy of the attosecond pulses after generation is estimated to be on the order of \SI{10}{pJ} by assuming a conversion efficiency of $10^{-6}$ as measured earlier \cite{harth_compact_2018}. The attosecond pulse duration is on the order of 400 to \SI{600}{as} (see supporting information). The attosecond pulse pair is collinearly overlapped with the remaining \SI{20}{\percent} of the NIR pulse not used for HHG with a drilled mirror, as described in more detail elsewhere \cite{mikaelsson_high-repetition_2020}. The delay between the NIR and XUV pulses is controlled using a piezo-driven stage and the NIR pulse energy of up to \SI{2.2}{\micro J} is adjusted via a motorized iris after recombination. An additional lens before the recombination mirror ensures that the NIR pulses have a similar wave front curvature as the attosecond pulse pair after generation \cite{wikmark_spatiotemporal_2019}. Two toroidal mirrors with focal lengths of $f_1=\SI{350}{mm}$ and $f_2=\SI{300}{mm}$, as described in \cite{mikaelsson_high-repetition_2020}, image the foci of the two collinear beams via an intermediate focus $4f_1$ after the initial focus onto the surface of a ZnO crystal at a distance of $4f_2$ to the intermediate focus and at an angle of \SI{65}{\degree} to the surface normal. Both the toroidal mirrors and the crystal are illuminated with p-polarized light. The incidence angle of light onto the flat ZnO crystal is close to the Brewster angle of the NIR wavelengths, resulting in a calculated reflectivity of only \SI{0.1}{\percent} over the whole spectral bandwidth. XUV light reflected from the ZnO crystal is used for beam diagnostics by directing it onto a microchannel plate detector equipped with a phosphor screen and a camera (see supporting information for more information concerning data treatment using beam diagnostics).

The ZnO crystal (MTI Corporation) is oriented in the (0001) direction relative to the surface and is polished to a roughness of \SI{0.5}{nm}. It is carefully prepared by repeated cycles of sputtering with argon ions and annealing up to $\sim$\SI{800}{\degree C} in an ultrahigh vacuum environment with a base pressure of less than \SI{5e-10}{mbar}. The cleanliness of the surface is confirmed by a separate measurement using a helium gas discharge light source (\SI{21.2}{eV}) and a hemispherical electron energy analyzer: A clear signature of the Zn-3d and O-2p states (Figure \ref{fig:Fig1}b,d)  is identified only after the cleaning process, indicating that residuals have been successfully removed from the surface. Binding energies of approximately \SI{4.3}{eV} and \SI{11.0}{eV} are measured for emission from the O-2p and Zn-3d states, respectively.

\begin{figure*}[h!tbp]
\centering
\includegraphics[width=\textwidth]{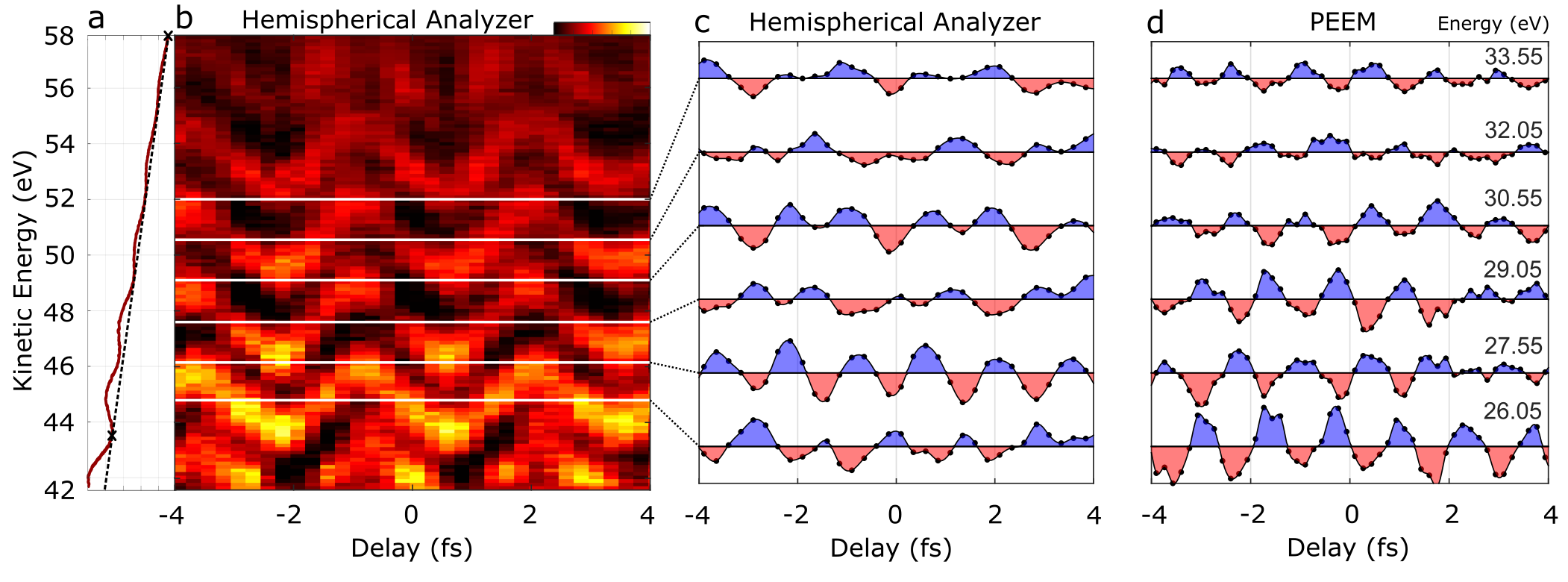}
\caption{Spectroscopy results from a ZnO surface. a) Kinetic energy spectrum of photoelectrons emitted from a clean ZnO surface by a pair of attosecond pulses. b) Kinetic energy spectra for different time delays between the attosecond pulse pair and an additional few-femtosecond infrared pulse. For clarity, a linear background, as indicated in a), is subtracted from the plotted spectra to make the observed periodic shifts of the peak positions better visible. The spectra seen for a wider range of time delays can be found in the SI. c) Lineouts of the dataset shown in b) for different kinetic energies spaced by the photon energy of the NIR pulse. Periodic oscillations with twice the NIR laser frequency and alternating phase between neighboring lineouts are observed. d) Lineouts produced from a comparable measurement, albeit at reduced kinetic energies, in a photoemission electron microscope showing very similar oscillations with alternating phase and twice the infrared laser frequency.}
\label{fig:Fig2}
\end{figure*}

When the attosecond pulse pair reaches the ZnO crystal, the high photon energy excites electrons from the Zn-3d and O-2p states into the continuum above the vacuum level energetically allowing them to escape the material.
The mean free paths of electrons in the energy range of 40 to \SI{60}{eV} is in the range of 0.5 to \SI{1}{nm} \cite{seah_quantitative_1979, shinotsuka_calculations_2019, flores-mancera_electron_2020} while the absorption length of the attosecond pulses is at least an order of magnitude longer.
Consequently, most of the electrons will inelastically scatter and subsequently lose their excess energy before they can reach the material-vacuum interface and leave the material.
As a result, only electrons excited from the first few atomic layers sufficiently close to the surface and with the k-vector pointing (primarily) in the direction of the surface normal, will reach the surface and leave the material (Figure \ref{fig:Fig1}a) resulting in an extremely surface sensitive probe.
The photoemitted electrons are detected using two different setups: a photoemission electron microscope (Focus IS-PEEM) or a hemispherical energy analyzer (Specs Phoibos 100).
In both cases, the kinetic energies of the photoelectrons are recorded.
The hemispherical analyzer is capable of simultaneous acquisition of electron spectra at different energies from the illuminated area of the sample while the PEEM acquires spectra in a serial manner using a high-pass filter but at the same time with spatial information about the electron emission site in the laser focus.

The attosecond XUV pulses illuminate an area of approximately \SI{50}{\micro m} x \SI{100}{\micro m} as shown in the inset of Figure \ref{fig:Fig1}c. The spot size of the co-propagating infrared pulses is adjusted to match the area illuminated by the attosecond pulses (see materials and methods). Thus, the photoelectrons after emission not only experience the static field by the sample near the surface but additionally the oscillating field of the focused NIR pulses. In this external field, electrons are accelerated and thus gain or lose kinetic energy depending on their emission time and direction. This change in the kinetic energy spectrum is recorded with either the hemisphere or the PEEM while the time delay between the attosecond XUV pulse pair and the infrared field is varied using a piezo-driven stage with sub-nanometer and thus attosecond precision.

\subsection{Attosecond interferometry with pairs of XUV pulses}

Without the infrared field, we record a periodically modulated kinetic energy spectrum which is shown in Figure \ref{fig:Fig2}a. The periodic modulation along the energy axis is a signature of the temporal structure of the XUV pulse pair and can be understood as follows: First, it is not only visible in the kinetic energy spectrum but also in the optical spectrum of the XUV pulse pair shown in Figure \ref{fig:Fig1}c. In a wave picture in the time domain, the two XUV pulses with a fixed temporal spacing interfere which leads to a sinusoidal amplitude modulation of the spectrum, similar to a double-slit experiment in the spatial domain.

However, in the experiment, we do not measure an optical, but an electron spectrum: If photoelectrons were emitted from a single, energetically well-defined state with energy-independent cross-section using this XUV pulse pair, the optical spectrum would be directly translated into a photoelectron spectrum, just shifted by the binding energy of the state and the detector work function. In our case, electrons are emitted from two different states (Zn-3d and O-2p) schematically shown in Figure \ref{fig:Fig1}b and measured in Figure \ref{fig:Fig1}d. The existence of such well-defined shallow core level states in the valence band region make ZnO a good material for this application. The energy difference of the two states of approximately \SI{6.7}{eV} coincides with twice the difference in energy between two consecutive peaks in the optical spectrum (Figure \ref{fig:Fig1}c and materials and methods for details). Thus, each peak observed in the electron spectrum is a superposition of emission from two different core levels. An exception are the highest kinetic energies where the O-2p state can be clearly identified as the origin of photoemission because the higher binding energy of the Zn-3d state prevents its electrons from reaching the same kinetic energies. More generally, the relative contribution of the two core levels to each peak in the electron spectrum depends both on the amplitude of the respective harmonic in the optical spectrum and the energy-dependent cross-section of the respective core level state \cite{piper_direct_2010}. This is taken into account in the approximation of the measured kinetic energy spectrum shown in Figure \ref{fig:Fig1}e, where the energy dependent cross-section for each of the two states is used as a fitting parameter to reproduce the data. More details on the fitting procedure can be found in the SI. 

Electron emission from comparably broad energy states with, compared to traditional experiments, broad distributions of photon energies (Figure \ref{fig:Fig1}d) leads to a substantial structure-less background. This is visualized in Figure \ref{fig:Fig1}e where the contributions from individual harmonics and states are shown. To better identify changes to the modulated part of the spectrum, a linear slope will be subtracted from the kinetic energy spectra in the following representations. It is calculated for the unperturbed reference spectrum as shown in Figure \ref{fig:Fig2}a and used in the following.

In Figure \ref{fig:Fig2}b, the slope-subtracted spectra are shown for varying delay between the attosecond pulse pair on the one hand and the NIR pulse on the other hand. A clear periodic shift of the maxima in the kinetic energy spectrum is observed. The frequency of the oscillation with varying delay corresponds to the laser frequency $\omega=2\pi c_0/\lambda$. This is similar to previous attosecond streaking experiments with isolated attosecond pulses in infrared fields both in the gas phase \cite{hentschel_attosecond_2001} and on solids \cite{cavalieri_attosecond_2007}. There, photoelectrons effectively gain or lose kinetic energy in the oscillating NIR field depending on their emission time relative to this field. An experiment with pairs of attosecond pulses as shown here can be even more sensitive to changes in the optical field because the mechanism leading to the periodic shift of the spectrum is slightly different \cite{cheng_controlling_2020}: When the first attosecond pulse emits, in a semi-classical picture, an electron at a time where the infrared field is, for example, leading to a net forward acceleration, the temporal distance of the two attosecond pulses ensures that an electron emitted by the second pulse will effectively lose kinetic energy. Thus, the two electron wave packets emitted by the two pulses will experience opposite phase changes in the temporal and spectral domain. In the case of isolated attosecond pulses, the change has to be sufficiently large to induce a measurable kinetic energy change. In our case of two attosecond pulses however, the interference between electron wave packets emitted during the first and second pulse makes already a linear spectral phase difference visible by an apparent shift in the peak positions. Also in this intermediate regime between isolated attosecond XUV pulses and pulse trains, a comparison of emission delays would be possible if the origin of photoemission was not obscured by the superposition of peaks discussed above \cite{gebauer_equivalence_2019}.

Instead, we want to compare the energy-resolved results obtained in a hemispherical analyzer with the PEEM which offers additional spatial resolution. For this purpose, lineouts of the signal at and in between the peak positions of the electron signal are calculated to account for the reduced spectral resolution of the PEEM used in the following. The energetic width used for averaging is equal to half the spacing of the peaks, i.e. all data is used in this process. This leads to the line plots in Figure \ref{fig:Fig2}c. Clear oscillations with $2\omega$, twice the laser frequency, are visible. The reason for this particular periodicity can be directly understood by looking at the dataset in Figure \ref{fig:Fig2}b: At energy positions between the peaks in the unperturbed spectrum, a maximum count rate is observed both when the peak at the next higher kinetic energy is shifted downwards and, half a temporal cycle later, when the lower peak is shifted upwards in energy. At the NIR field strength used in the experiment, the observed periodicity of $2\omega$ is independent of the central energy used for energetic averaging. Looking at different kinetic energies, only a negligible phase shift of the oscillations is observed. This suggests a temporally well-compressed attosecond pulse structure.

\begin{figure*}[h!tbp]
\centering
\includegraphics[width=\textwidth]{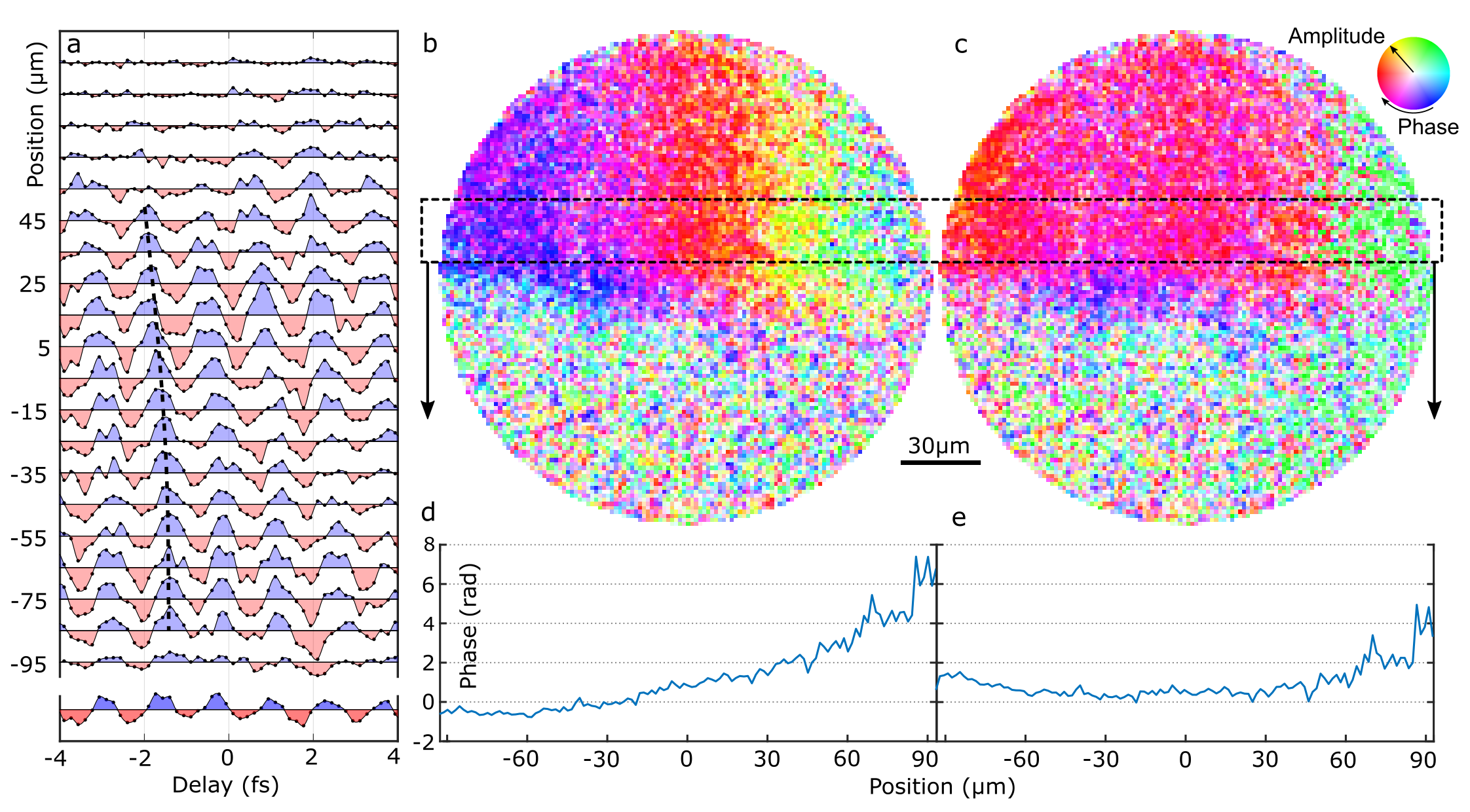}
\caption{Spatially resolved attosecond interferometry on a ZnO surface.
a) Electron count rates for varying delay between NIR and XUV at different positions on the surface. Instead of averaging over both spatial dimensions of the surface as in Figure \ref{fig:Fig2}d, only the vertical direction is used for averaging. Data for a kinetic energy of \SI{26.05}{eV} are shown. A clear shift of the oscillation phase with the horizontal position can be observed (dashed line as a guide to the eye). The bottom most plot shows the sum of the measurements above (divided by 10) and resembles the measurement shown in the bottom of Figure \ref{fig:Fig2}d.
b) Spatial map of the amplitude and phase of the periodic oscillations shown in Figure \ref{fig:Fig2}d or \ref{fig:Fig3}a, but with higher spatial resolution (\SI{0.75}{\micro m} per square pixel). While the hue indicates the phase in a range from $0$ to $2\pi$, the saturation indicates the amplitude of the oscillations. As expected, the signal amplitude follows the homogeneous spatial distribution of the NIR field (see materials and methods) and not the significantly more inhomogeneous distribution of the XUV field (Figure \ref{fig:Fig1}c). A horizontal lineout of the phase averaged over the region marked by the dashed rectangle is shown in part d) of the Figure.
c) The same dataset as in a) but with a linear phase subtracted in horizontal direction to account for a potential misalignment of the two beams in this direction. The corresponding lineout is shown in part e).}
\label{fig:Fig3}
\end{figure*}

Now, we compare this to results we obtained in the PEEM. To make the data comparable, we first need to average the measurement signal over the simultaneously recorded spatial domain which is not available in the data recorded with the hemispherical analyzer. Due to limitations in the way kinetic energy spectra are recorded, the PEEM data show a reduced energy resolution. Thus, we compare spectra from the PEEM (Figure \ref{fig:Fig2}d) with those we obtained after averaging over parts of the kinetic spectrum recorded with the hemisphere (Figure \ref{fig:Fig2}c) to match the spectral resolution. The PEEM data (plotted in Figure \ref{fig:Fig2}d), which was recorded at slightly reduced kinetic energy of the electrons, shows oscillations with the same periodicity as data recorded with the hemisphere. In both cases, the phase of this oscillation alternates by approximately \SI{180}{\degree} between adjacent recorded energies, separated by the photon energy of the NIR. This directly indicates that also with the PEEM, despite the lower energy resolution, the same attosecond electron wave packet interference is observed as in the hemisphere experiment discussed before.

This is the first time a XUV attosecond time-resolved experiment has been demonstrated in a PEEM. The experiment is made possible by avoiding the regime of multiple electrons emitted per laser shot and hence space charge related reduction of energy resolution in the microscope column. The repetition rate of \SI{200}{kHz} enables the acquisition of sufficient statistics while keeping the number of electrons emitted per laser shot below 1. Additionally, ZnO lends itself well for such an experiment because it offers, after careful preparation, well-defined states for photoemission, giving clear signatures in the kinetic energy spectrum that can be tracked to obtain a temporally modulated signal. Finally, the use of pairs of XUV pulses creates a more structured kinetic energy spectrum compared to isolated pulses while it maintains the high temporal resolution. Changes to the feature-rich spectrum are more easily extracted in a measurement as shown here. 

\subsection{Spatially resolved attosecond interferometry}
As a last step, we look at the phase of the previously found oscillations using the spatial resolution of our microscope. Since the sample consists of a flat ZnO crystal with no resolvable features, no influence by the lateral structure of the sample is expected. Instead, this gives us the opportunity to spatially resolve the temporal structure of the attosecond XUV pulse pair relative to the phase of the infrared laser field. This is an important first step for any future attosecond time-resolved experiments on heterogeneous samples. Without this information, dynamics on the sample and inhomogeneities in the pulse structure could not be distinguished.

To visualize the multidimensional dataset, we need to compress the information. We average over the vertical dimension and use a bin size of \SI{5}{\micro m} in the horizontal direction (i.e. the direction of laser pulse propagation relative to the surface, shown in Figure \ref{fig:Fig1}a). This results in essentially the same curves as in Figure \ref{fig:Fig2}d, but now additionally for each horizontal bin (and each of the six energies as before). For simplicity, we only plot the lowest energy data in Figure \ref{fig:Fig3}a. A clear shift of the $2\omega$ oscillation along the horizontal position becomes visible. As described before, the modulation results from the electrons emitted by the XUV pulse pair interacting with the NIR field. A spatial phase shift of this modulation can originate from either internal differences of the sample or external differences by the applied optical fields. Because we probe a flat, feature-less surface, we believe that the shift originates from the pulses themselves and is not related to the properties of the ZnO crystal.

The external origin of the observed phase difference could be attributed to the NIR field interacting with the photoelectrons after emission or with the electrons while they are still below the surface. We illuminate the sample at an angle of \SI{65}{\degree} to the surface normal in horizontal direction. Hence, parts at larger position values in Figure \ref{fig:Fig3}a are illuminated before those at smaller values. However, the speed of light gives an upper limit for the interaction of different parts of the sample, ruling out the possibility of a part on the right of the sample (which is illuminated first) influencing a part on the left (which is illuminated last) on the time scale observed here as information requires more than \SI{300}{fs} to propagate across the observed sample area. The asynchronous illumination of different sample positions due to the gracing illumination is only insignificantly modified by height differences of the sample on the atomic scale.

The excitation intensity of the sample is clearly inhomogeneous. This could have a temporal influence on the phase of the light-electron interaction, because locally excited charges might alter the field at the surface. A local extreme in the measured phase would be expected at the infrared field's peak position at the center of the observed area which might have some contribution to the measurement, but cannot explain the dominant linear phase change across the sample. Hence, we conclude that the main origin for the observed phase change has to be found outside the material in the interaction with the infrared field, i.e. after emission of photoelectrons from the surface by the XUV pulse pair.

To investigate the interaction of photoelectrons with the infrared field, we reduce the complexity along the time axis and further increase the spatial resolution. We do this by calculating the amplitude $A_n$ and phase $\varphi_n$ of the observed $2\omega$ oscillation at each position and each of the six measured energy steps $E_n$ (see Figure \ref{fig:Fig2}d). Thus, we compress the temporal information into two values $A_n$ and $\varphi_n$ for each position and each energy. Finally, we average amplitude and phase over the energy dimension $\sum_n{A_n\cdot\exp{(i\varphi_n+n\pi)}}$, taking the alternating phase between adjacent energies into account. The result is shown in Figure \ref{fig:Fig3}b for each position in an amplitude and phase plot. Color indicates the phase while the saturation of the color indicates the relative amplitude of the measured oscillation.

It becomes apparent that there is a clear phase change across the measured area of the laser focus. The most significant feature is a linear phase gradient of the $2\omega$ oscillation in horizontal direction with a total phase change of approximately $\pi$ or a phase gradient of approximately $\pi/\SI{150}{\micro m}$ in this grazing incidence geometry. Additionally, there are smaller deviations from this trend at the top and in the middle of the image, indicating a more complex temporal relationship between the attosecond pulse pair envelope and the phase of the NIR pulse.

We found already that the temporal relation between the NIR pulse and the XUV pulse pair has to be responsible for the observed phase change of the oscillations. By estimating the Rayleigh length of the NIR beam, we find that the Gouy phase shift across the NIR focus can only marginally contribute to the experimentally observed phase shift. What remains is a potential angle $\alpha=\SI{3.4}{mrad}$ between the NIR and the XUV beam in horizontal direction which might cause the linear difference in phase observed in the experiment (see supporting information for a detailed derivation).

A quadratic phase term of approximately \SI{0.2}{mrad/\micro\meter\squared} remains in horizontal direction after subtraction of the linear part discussed before in light of an angular misalignment. This is shown in Figure \ref{fig:Fig3}c,e and a misalignment of the focal spot positions of the two beams in propagation direction, parallel to the optical axis, might be the cause of this, leading to a mismatch of the involved wavefronts. Looking closer at Figure \ref{fig:Fig3}c, one finds that the phase at the lower end of the measured region seems to be phase shifted by approximately $\pi/2$. Surface imperfections of the different optics guiding the NIR and XUV beams might be the cause of this, indicated already by the structured beam profile seen in the inset of Figure \ref{fig:Fig1}c.


\section{Conclusions}
Utilizing a ZnO surface, we have performed a time-resolved attosecond interferometry experiment which spatially resolved the participating light foci. We showed that such an experiment is feasible using a high repetition rate laser system and a sample with well-defined energy states such that a broad attosecond pulse spectrum can be used. We observe a mostly linear phase variation in the attosecond interferometry signal along the projection of the light path on the surface, indicating a slight angular misalignment of the NIR and attosecond pulse beam relative to each other by less than \SI{3.4}{mrad}. More subtle phase differences with a more complicated spatial dependency are attributed to actual wave front differences between the XUV and the NIR beam on the sample surface.

The demonstrated high sensitivity to phase differences of optical electric fields atomically close to a material vacuum interface is essential for the improvement of opto-electronic devices or the enhancement of photocatalysis as the transient dynamics at such interfaces define the resulting macroscopic function. 

The spatial phase differences measured in the PEEM experiment (Figure \ref{fig:Fig3}b) point towards a spatial resolution on the order of 1 to \SI{10}{\micro m}, however there is no hindrance for observing changes with the maximum spatial resolution of the PEEM on the few tens of nanometer scale. Main limitation would be the electron count rate in smaller ares, which can be counteracted by reducing the light spot sizes together with the field of view of the PEEM. This makes dynamic studies with the highest resolution conceivable, significantly better than achievable optical focus sizes both in the visible and XUV spectral range \cite{motoyama_broadband_2019}.

In future experiments, utilizing a time-of-flight spectrometer for improved statistics \cite{lin_time_2009}, the observed spatial phase difference needs to be carefully characterized before conclusions about actual dynamics happening on a heterogeneous surface can be drawn. Tighter focusing of the involved light pulses to match the desired field-of-view while keeping the intensities constant, as it is required to avoid detrimental space charge effects when investigating smaller areas, does not alleviate the challenges observed here as they are coupled to the focal spot size. To minimize the phase differences, matching the curvature of both the XUV and the NIR wave fronts by choosing the same focusing conditions seems essential.


\section{Materials and methods}

\subsection{Details on pulses and beam paths}
\begin{figure*}[h!tbp]
\centering
\includegraphics[width=0.4\textwidth]{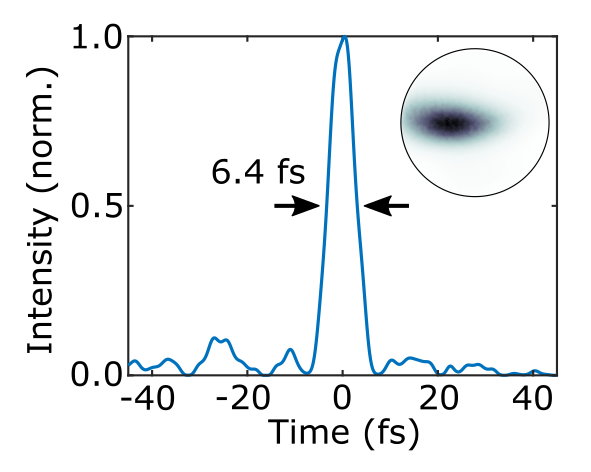}
\caption{Temporal structure (intensity envelope) of the few-cycle NIR laser pulses characterized using the dispersion scan technique. The inset shows the beam profile recorded in the PEEM which was illuminated at an angle of \SI{65}{\degree} to the surface normal and emitted photoelectrons in a nonlinear multiphoton photoemission process ($n=4$).}
\label{fig:FigPulseDuration}
\end{figure*}

The pulse duration of the NIR pulses was measured using the dispersion scan technique \cite{miranda_characterization_2012,sytcevich_characterizing_2021}. The resulting intensity envelope of the pulses is shown in Figure \ref{fig:FigPulseDuration}. The carrier to envelope phase is not measured using this technique. Instead, it was adjusted while observing the attosecond interferometry result (Figure \ref{fig:Fig2}b).

Two toroidal mirrors with focal lengths of $f_1=\SI{350}{mm}$ and $f_2=\SI{300}{mm}$ image the focus of the XUV generation via an intermediate focus $4f_1$ after the initial focus onto the surface of a ZnO crystal at a distance of $4f_2$ to the intermediate focus and at an angle of \SI{65}{\degree} to the surface normal. The spot size of the XUV beam is characterized by imaging photoelectrons emitted from the surface via a single photon photoemission process. The PEEM including the ZnO crystal is positioned such that the minimum XUV spot size is reached on the crystal. 

The NIR light used in the experiment is focused close to the XUV generation before it is collinearly recombined with the XUV light using a drilled mirror. The focus position of the NIR beam and the angle of the recombination mirror is adjusted to match the NIR focus position to the XUV focus position.

The size of the NIR focus is adjusted using a motorized iris aperture placed between the NIR focus and the first toroidal mirror. It simultaneously adjusts the NIR intensity reached on the sample surface. As the NIR beam has a donut like shape after the recombination mirror with the XUV beam propagating in the center, the iris does not affect the XUV light.

\subsection{Calculation of kinetic energies}
Electrons originating from different states are emitted from the ZnO surface after absorption of different photon energies. As an example, an electron detected with a kinetic energy of \SI{53.5}{eV} was potentially photoemitted by a photon with an energy of \SI{62.1}{eV} from the O-2p state (37\textsuperscript{th} harmonic; \SI{4.3}{eV} binding energy plus \SI{4.3}{eV} detector work function) or by a photon with an energy of \SI{68.8}{eV} (41\textsuperscript{st} harmonic, \SI{11.0}{eV} binding energy plus \SI{4.3}{eV} detector work function) from the Zn-3d state.

This coincidence of the difference in work function with the difference between spectral peaks of the ionizing radiation makes the electrons’ origin indistinguishable in the kinetic energy spectrum. Only at the highest kinetic energies the O-2p state can be clearly identified as the origin of photoemission because the higher binding energy of the Zn-3d state prevents its electrons from reaching the same kinetic energies. Still, the higher cross section of the Zn-3d compared to the O-2p state makes it challenging to separate the two.

\section{Acknowledgments} The authors would like to thank Christoph Lienau for the opportunity to utilize the hemispherical analyzer. This research was financially supported by the Vetenskapsrådet (2013-8185, 2014-04580, 2016-04907, 2020-04201), the European Commission (793604 ATTOPIE), the European Research Council (884900), the Knut och Alice Wallenbergs Stiftelse, the Wallenberg Center for Quantum Technology, and Laserlab-Europe (654148).


\section{Conflict of interest} The authors declare no competing interests.

\section{Data availability} Data underlying the results presented in this paper are not publicly available at this time but may be obtained from the authors upon reasonable request.

\section{Keywords}
time-resolved photoemission electron microscopy, attosecond interferometry, zinc oxide, surface science

\section{References}

\bibliography{janszotero}

\begin{thebibliography}{10}
\providecommand{\url}[1]{\texttt{#1}}
\providecommand{\urlprefix}{URL }

\bibitem{sainadh_attosecond_2019}
U.~S. Sainadh, H.~Xu, X.~Wang, A.~Atia-Tul-Noor, W.~C. Wallace, N.~Douguet,
  A.~Bray, I.~Ivanov, K.~Bartschat, A.~Kheifets, R.~T. Sang, I.~V. Litvinyuk.
\newblock \emph{Nature} \textbf{2019}, \emph{568}, 7750 75.

\bibitem{nisoli_attosecond_2017}
M.~Nisoli, P.~Decleva, F.~Calegari, A.~Palacios, F.~Martín.
\newblock \emph{Chemical Reviews} \textbf{2017}, \emph{117}, 16 10760.

\bibitem{jordan_attosecond_2020}
I.~Jordan, M.~Huppert, D.~Rattenbacher, M.~Peper, D.~Jelovina, C.~Perry, A.~von
  Conta, A.~Schild, H.~J. Wörner.
\newblock \emph{Science} \textbf{2020}, \emph{369}, 6506 974.

\bibitem{cavalieri_attosecond_2007}
A.~L. Cavalieri, N.~Müller, T.~Uphues, V.~S. Yakovlev, A.~Baltuška,
  B.~Horvath, B.~Schmidt, L.~Blümel, R.~Holzwarth, S.~Hendel, M.~Drescher,
  U.~Kleineberg, P.~M. Echenique, R.~Kienberger, F.~Krausz, U.~Heinzmann.
\newblock \emph{Nature} \textbf{2007}, \emph{449}, 7165 1029.

\bibitem{ossiander_speed_2022}
M.~Ossiander, K.~Golyari, K.~Scharl, L.~Lehnert, F.~Siegrist, J.~P. Bürger,
  D.~Zimin, J.~A. Gessner, M.~Weidman, I.~Floss, V.~Smejkal, S.~Donsa,
  C.~Lemell, F.~Libisch, N.~Karpowicz, J.~Burgdörfer, F.~Krausz, M.~Schultze.
\newblock \emph{Nature Communications} \textbf{2022}, \emph{13}, 1 1620.

\bibitem{sederberg_attosecond_2020}
S.~Sederberg, D.~Zimin, S.~Keiber, F.~Siegrist, M.~S. Wismer, V.~S. Yakovlev,
  I.~Floss, C.~Lemell, J.~Burgdörfer, M.~Schultze, F.~Krausz, N.~Karpowicz.
\newblock \emph{Nature Communications} \textbf{2020}, \emph{11}, 1 430.

\bibitem{bionta_-chip_2021}
M.~R. Bionta, F.~Ritzkowsky, M.~Turchetti, Y.~Yang, D.~Cattozzo~Mor, W.~P.
  Putnam, F.~X. Kärtner, K.~K. Berggren, P.~D. Keathley.
\newblock \emph{Nature Photonics} \textbf{2021}, \emph{15}, 6 456.

\bibitem{sommer_attosecond_2016}
A.~Sommer, E.~M. Bothschafter, S.~A. Sato, C.~Jakubeit, T.~Latka,
  O.~Razskazovskaya, H.~Fattahi, M.~Jobst, W.~Schweinberger, V.~Shirvanyan,
  V.~S. Yakovlev, R.~Kienberger, K.~Yabana, N.~Karpowicz, M.~Schultze,
  F.~Krausz.
\newblock \emph{Nature} \textbf{2016}, \emph{534}, 7605 86.

\bibitem{vampa_linking_2015}
G.~Vampa, T.~J. Hammond, N.~Thiré, B.~E. Schmidt, F.~Légaré, C.~R. McDonald,
  T.~Brabec, P.~B. Corkum.
\newblock \emph{Nature} \textbf{2015}, \emph{522}, 7557 462.

\bibitem{luu_extreme_2015}
T.~T. Luu, M.~Garg, S.~Y. Kruchinin, A.~Moulet, M.~T. Hassan, E.~Goulielmakis.
\newblock \emph{Nature} \textbf{2015}, \emph{521}, 7553 498.

\bibitem{goulielmakis_high_2022}
E.~Goulielmakis, T.~Brabec.
\newblock \emph{Nature Photonics} \textbf{2022}, \emph{16}, 6 411.

\bibitem{lucchini_attosecond_2016}
M.~Lucchini, S.~A. Sato, A.~Ludwig, J.~Herrmann, M.~Volkov, L.~Kasmi,
  Y.~Shinohara, K.~Yabana, L.~Gallmann, U.~Keller.
\newblock \emph{Science} \textbf{2016}, \emph{353}, 6302 916.

\bibitem{schultze_attosecond_2014}
M.~Schultze, K.~Ramasesha, C.~D. Pemmaraju, S.~A. Sato, D.~Whitmore,
  A.~Gandman, J.~S. Prell, L.~J. Borja, D.~Prendergast, K.~Yabana, D.~M.
  Neumark, S.~R. Leone.
\newblock \emph{Science} \textbf{2014}, \emph{346}, 6215 1348.

\bibitem{kobayashi_direct_2019}
Y.~Kobayashi, K.~F. Chang, T.~Zeng, D.~M. Neumark, S.~R. Leone.
\newblock \emph{Science} \textbf{2019}, \emph{365}, 6448 79.

\bibitem{tao_direct_2016}
Z.~Tao, C.~Chen, T.~Szilvasi, M.~Keller, M.~Mavrikakis, H.~Kapteyn, M.~Murnane.
\newblock \emph{Science} \textbf{2016}, \emph{353}, 6294 62.

\bibitem{chen_distinguishing_2017}
C.~Chen, Z.~Tao, A.~Carr, P.~Matyba, T.~Szilvási, S.~Emmerich, M.~Piecuch,
  M.~Keller, D.~Zusin, S.~Eich, M.~Rollinger, W.~You, S.~Mathias, U.~Thumm,
  M.~Mavrikakis, M.~Aeschlimann, P.~M. Oppeneer, H.~Kapteyn, M.~Murnane.
\newblock \emph{Proceedings of the National Academy of Sciences} \textbf{2017},
  \emph{114}, 27 E5300.

\bibitem{siek_angular_2017}
F.~Siek, S.~Neb, P.~Bartz, M.~Hensen, C.~Strüber, S.~Fiechter,
  M.~Torrent-Sucarrat, V.~M. Silkin, E.~E. Krasovskii, N.~M. Kabachnik,
  S.~Fritzsche, R.~D. Muiño, P.~M. Echenique, A.~K. Kazansky, N.~Müller,
  W.~Pfeiffer, U.~Heinzmann.
\newblock \emph{Science} \textbf{2017}, \emph{357}, 6357 1274.

\bibitem{spektor_revealing_2017}
G.~Spektor, D.~Kilbane, A.~K. Mahro, B.~Frank, S.~Ristok, L.~Gal, P.~Kahl,
  D.~Podbiel, S.~Mathias, H.~Giessen, F.-J. M.~z. Heringdorf, M.~Orenstein,
  M.~Aeschlimann.
\newblock \emph{Science} \textbf{2017}, \emph{355}, 6330 1187.

\bibitem{podbiel_imaging_2017}
D.~Podbiel, P.~Kahl, A.~Makris, B.~Frank, S.~Sindermann, T.~J. Davis,
  H.~Giessen, M.~H.-v. Hoegen, F.-J. Meyer~zu Heringdorf.
\newblock \emph{Nano Letters} \textbf{2017}, \emph{17}, 11 6569, publisher:
  American Chemical Society.

\bibitem{dai_plasmonic_2020}
Y.~Dai, Z.~Zhou, A.~Ghosh, R.~S.~K. Mong, A.~Kubo, C.-B. Huang, H.~Petek.
\newblock \emph{Nature} \textbf{2020}, \emph{588}, 7839 616.

\bibitem{zhong_nonlinear_2020}
J.-H. Zhong, J.~Vogelsang, J.-M. Yi, D.~Wang, L.~Wittenbecher, S.~Mikaelsson,
  A.~Korte, A.~Chimeh, C.~L. Arnold, P.~Schaaf, E.~Runge, A.~L. Huillier,
  A.~Mikkelsen, C.~Lienau.
\newblock \emph{Nature Communications} \textbf{2020}, \emph{11}, 1 1464.

\bibitem{wittenbecher_unraveling_2021}
L.~Wittenbecher, E.~Viñas~Boström, J.~Vogelsang, S.~Lehman, K.~A. Dick,
  C.~Verdozzi, D.~Zigmantas, A.~Mikkelsen.
\newblock \emph{ACS Nano} \textbf{2021}, \emph{15}, 1 1133, publisher: American
  Chemical Society.

\bibitem{stockman_attosecond_2007}
M.~I. Stockman, M.~F. Kling, U.~Kleineberg, F.~Krausz.
\newblock \emph{Nature Photonics} \textbf{2007}, \emph{1}, 9 539.

\bibitem{mikkelsen_photoemission_2009}
A.~Mikkelsen, J.~Schwenke, T.~Fordell, G.~Luo, K.~Klünder, E.~Hilner,
  N.~Anttu, A.~A. Zakharov, E.~Lundgren, J.~Mauritsson, J.~N. Andersen, H.~Q.
  Xu, A.~L’Huillier.
\newblock \emph{Review of Scientific Instruments} \textbf{2009}, \emph{80}, 12
  123703.

\bibitem{gates_new_2005}
B.~D. Gates, Q.~Xu, M.~Stewart, D.~Ryan, C.~G. Willson, G.~M. Whitesides.
\newblock \emph{Chemical Reviews} \textbf{2005}, \emph{105}, 4 1171.

\bibitem{hengsteler_bringing_2021}
J.~Hengsteler, B.~Mandal, C.~van Nisselroy, G.~P.~S. Lau, T.~Schlotter,
  T.~Zambelli, D.~Momotenko.
\newblock \emph{Nano Letters} \textbf{2021}, \emph{21}, 21 9093.

\bibitem{aeschlimann_adaptive_2007}
M.~Aeschlimann, M.~Bauer, D.~Bayer, T.~Brixner, F.~J. García~de Abajo,
  W.~Pfeiffer, M.~Rohmer, C.~Spindler, F.~Steeb.
\newblock \emph{Nature} \textbf{2007}, \emph{446}, 7133 301.

\bibitem{novotny_principles_2006}
L.~Novotny, B.~Hecht.
\newblock \emph{Principles of {Nano}-{Optics}}.
\newblock Cambridge University Press, \textbf{2006}.

\bibitem{boolakee_light-field_2022}
T.~Boolakee, C.~Heide, A.~Garzón-Ramírez, H.~B. Weber, I.~Franco,
  P.~Hommelhoff.
\newblock \emph{Nature} \textbf{2022}, \emph{605}, 7909 251, number: 7909
  Publisher: Nature Publishing Group.

\bibitem{vogelsang_coherent_2021}
J.~Vogelsang, L.~Wittenbecher, D.~Pan, J.~Sun, S.~Mikaelsson, C.~L. Arnold,
  A.~L’Huillier, H.~Xu, A.~Mikkelsen.
\newblock \emph{ACS Photonics} \textbf{2021}, \emph{8}, 6 1607.

\bibitem{feist_ultrafast_2017}
A.~Feist, N.~Bach, N.~Rubiano~da Silva, T.~Danz, M.~Möller, K.~E. Priebe,
  T.~Domröse, J.~G. Gatzmann, S.~Rost, J.~Schauss, S.~Strauch, R.~Bormann,
  M.~Sivis, S.~Schäfer, C.~Ropers.
\newblock \emph{Ultramicroscopy} \textbf{2017}, \emph{176} 63.

\bibitem{garg_real-space_2021}
M.~Garg, A.~Martin-Jimenez, M.~Pisarra, Y.~Luo, F.~Martín, K.~Kern.
\newblock \emph{Nature Photonics} \textbf{2021}, 1--7.

\bibitem{cocker_tracking_2016}
T.~L. Cocker, D.~Peller, P.~Yu, J.~Repp, R.~Huber.
\newblock \emph{Nature} \textbf{2016}, \emph{539}, 7628 263.

\bibitem{eisele_ultrafast_2014}
M.~Eisele, T.~L. Cocker, M.~A. Huber, M.~Plankl, L.~Viti, D.~Ercolani,
  L.~Sorba, M.~S. Vitiello, R.~Huber.
\newblock \emph{Nature Photonics} \textbf{2014}, \emph{8} 841.

\bibitem{gaida_attosecond_2023}
J.~H. Gaida, H.~Lourenço-Martins, M.~Sivis, T.~Rittmann, A.~Feist, F.~J.~G.
  de~Abajo, C.~Ropers.
\newblock Attosecond electron microscopy by free-electron homodyne detection,
  \textbf{2023}.
\newblock \urlprefix\url{http://arxiv.org/abs/2305.03005}.
\newblock ArXiv:2305.03005 [cond-mat, physics:physics, physics:quant-ph].

\bibitem{nabben_attosecond_2023}
D.~Nabben, J.~Kuttruff, L.~Stolz, A.~Ryabov, P.~Baum.
\newblock \emph{Nature} \textbf{2023}, 1--5, publisher: Nature Publishing
  Group.

\bibitem{forg_attosecond_2016}
B.~Förg, J.~Schötz, F.~Süßmann, M.~Förster, M.~Krüger, B.~Ahn, W.~A.
  Okell, K.~Wintersperger, S.~Zherebtsov, A.~Guggenmos, V.~Pervak, A.~Kessel,
  S.~A. Trushin, A.~M. Azzeer, M.~I. Stockman, D.~Kim, F.~Krausz,
  P.~Hommelhoff, M.~F. Kling.
\newblock \emph{Nature Communications} \textbf{2016}, \emph{7} 11717.

\bibitem{zherebtsov_controlled_2011}
S.~Zherebtsov, T.~Fennel, J.~Plenge, E.~Antonsson, I.~Znakovskaya, A.~Wirth,
  O.~Herrwerth, F.~Süßmann, C.~Peltz, I.~Ahmad, S.~A. Trushin, V.~Pervak,
  S.~Karsch, M.~J.~J. Vrakking, B.~Langer, C.~Graf, M.~I. Stockman, F.~Krausz,
  E.~Rühl, M.~F. Kling.
\newblock \emph{Nature Physics} \textbf{2011}, \emph{7}, 8 656.

\bibitem{seiffert_strong-field_2022}
L.~Seiffert, S.~Zherebtsov, M.~F. Kling, T.~Fennel.
\newblock \emph{Advances in Physics: X} \textbf{2022}, \emph{7}, 1 2010595.

\bibitem{hoff_tracing_2017}
D.~Hoff, M.~Krüger, L.~Maisenbacher, A.~M. Sayler, G.~G. Paulus,
  P.~Hommelhoff.
\newblock \emph{Nature Physics} \textbf{2017}, \emph{13}, 10 947.

\bibitem{wikmark_spatiotemporal_2019}
H.~Wikmark, C.~Guo, J.~Vogelsang, P.~W. Smorenburg, H.~Coudert-Alteirac,
  J.~Lahl, J.~Peschel, P.~Rudawski, H.~Dacasa, S.~Carlström, S.~Maclot, M.~B.
  Gaarde, P.~Johnsson, C.~L. Arnold, A.~L’Huillier.
\newblock \emph{Proceedings of the National Academy of Sciences} \textbf{2019},
  \emph{116}, 11 4779.

\bibitem{vincenti_attosecond_2012}
H.~Vincenti, F.~Quéré.
\newblock \emph{Physical Review Letters} \textbf{2012}, \emph{108}, 11 113904.

\bibitem{lee_wave-front_2003}
D.~G. Lee, J.~J. Park, J.~H. Sung, C.~H. Nam.
\newblock \emph{Optics Letters} \textbf{2003}, \emph{28}, 6 480.

\bibitem{dacasa_single-shot_2019}
H.~Dacasa, H.~Coudert-Alteirac, C.~Guo, E.~Kueny, F.~Campi, J.~Lahl,
  J.~Peschel, H.~Wikmark, B.~Major, E.~Malm, D.~Alj, K.~Varjú, C.~L. Arnold,
  G.~Dovillaire, P.~Johnsson, A.~L’Huillier, S.~Maclot, P.~Rudawski,
  P.~Zeitoun.
\newblock \emph{Optics Express} \textbf{2019}, \emph{27}, 3 2656.

\bibitem{lin_time_2009}
J.~Lin, N.~Weber, A.~Wirth, S.~H. Chew, M.~Escher, M.~Merkel, M.~F. Kling,
  M.~I. Stockman, F.~Krausz, U.~Kleineberg.
\newblock \emph{Journal of Physics: Condensed Matter} \textbf{2009}, \emph{21},
  31 314005.

\bibitem{mikaelsson_high-repetition_2020}
S.~Mikaelsson, J.~Vogelsang, C.~Guo, I.~Sytcevich, A.-L. Viotti, F.~Langer,
  Y.-C. Cheng, S.~Nandi, W.~Jin, A.~Olofsson, R.~Weissenbilder, J.~Mauritsson,
  A.~L’Huillier, M.~Gisselbrecht, C.~L. Arnold.
\newblock \emph{Nanophotonics} \textbf{2020}, \emph{10}, 1 117.

\bibitem{harth_compact_2018}
A.~Harth, C.~Guo, Y.-C. Cheng, A.~Losquin, M.~Miranda, S.~Mikaelsson, C.~M.
  Heyl, O.~Prochnow, J.~Ahrens, U.~Morgner, A.~L’Huillier, C.~L. Arnold.
\newblock \emph{Journal of Optics} \textbf{2018}, \emph{20}, 1 014007.

\bibitem{seah_quantitative_1979}
M.~P. Seah, W.~A. Dench.
\newblock \emph{Surface and interface analysis} \textbf{1979}, \emph{1}, 1 2.

\bibitem{shinotsuka_calculations_2019}
H.~Shinotsuka, S.~Tanuma, C.~J. Powell, D.~R. Penn.
\newblock \emph{Surface and Interface Analysis} \textbf{2019}, \emph{51}, 4
  427.

\bibitem{flores-mancera_electron_2020}
M.~A. Flores-Mancera, J.~S. Villarrubia, G.~Massillon-JL.
\newblock \emph{ACS Omega} \textbf{2020}, \emph{5}, 8 4139.

\bibitem{piper_direct_2010}
L.~F.~J. Piper, A.~R.~H. Preston, A.~Fedorov, S.~W. Cho, A.~DeMasi, K.~E.
  Smith.
\newblock \emph{Physical Review B} \textbf{2010}, \emph{81}, 23 233305.

\bibitem{hentschel_attosecond_2001}
M.~Hentschel, R.~Kienberger, C.~Spielmann, G.~A. Reider, N.~Milosevic,
  T.~Brabec, P.~Corkum, U.~Heinzmann, M.~Drescher, F.~Krausz.
\newblock \emph{Nature} \textbf{2001}, \emph{414}, 6863 509.

\bibitem{cheng_controlling_2020}
Y.-C. Cheng, S.~Mikaelsson, S.~Nandi, L.~Rämisch, C.~Guo, S.~Carlström,
  A.~Harth, J.~Vogelsang, M.~Miranda, C.~L. Arnold, A.~L’Huillier,
  M.~Gisselbrecht.
\newblock \emph{Proceedings of the National Academy of Sciences} \textbf{2020},
  \emph{117}, 20 10727.

\bibitem{gebauer_equivalence_2019}
A.~Gebauer, S.~Neb, W.~Enns, B.~Stadtmüller, M.~Aeschlimann, W.~Pfeiffer.
\newblock \emph{Applied Sciences} \textbf{2019}, \emph{9}, 3 592.

\bibitem{motoyama_broadband_2019}
H.~Motoyama, A.~Iwasaki, Y.~Takei, T.~Kume, S.~Egawa, T.~Sato, K.~Yamanouchi,
  H.~Mimura.
\newblock \emph{Applied Physics Letters} \textbf{2019}, \emph{114}, 24 241102.

\bibitem{miranda_characterization_2012}
M.~Miranda, C.~L. Arnold, T.~Fordell, F.~Silva, B.~Alonso, R.~Weigand,
  A.~L’Huillier, H.~Crespo.
\newblock \emph{Optics Express} \textbf{2012}, \emph{20}, 17 18732.

\bibitem{sytcevich_characterizing_2021}
I.~Sytcevich, C.~Guo, S.~Mikaelsson, J.~Vogelsang, A.-L. Viotti, B.~Alonso,
  R.~Romero, P.~T. Guerreiro, I.~J. Sola, A.~L’Huillier, H.~Crespo,
  M.~Miranda, C.~L. Arnold.
\newblock \emph{Journal of the Optical Society of America B} \textbf{2021},
  \emph{38}, 5 1546.

\end{thebibliography}
\bibliographystyle{advancedmaterials}

\end{document}